# Are physiological oscillations *physiological*?

Lingyun (Ivy) Xiong[1,2] *, Alan Garfinkel[3]

1. Department of Stem Cell Biology and Regenerative Medicine, Eli and Edythe Broad Center for Regenerative Medicine and Stem Cell Research, Keck School of Medicine of the University of Southern California, Los Angeles, CA 90089, USA
2. Department of Quantitative and Computational Biology, University of Southern California, Los Angeles, CA 90089 USA
3. Departments of Medicine (Cardiology) and Integrative Biology and Physiology, University of California, Los Angeles, CA 90095 USA

*Correspondence: lingyunx@usc.edu

## Abstract

Despite widespread and striking examples of physiological oscillations, their functional role is often unclear. Even glycolysis, the paradigm example of oscillatory biochemistry, has seen questions about its oscillatory function. Here, we take a systems approach to summarize evidence that oscillations play critical physiological roles. Oscillatory behavior enables systems to avoid desensitization, to avoid chronically high and therefore toxic levels of chemicals, and to become more resistant to noise. Oscillation also enables complex physiological systems to reconcile incompatible conditions such as oxidation and reduction, by cycling between them, and to synchronize the oscillations of many small units into one large effect. In pancreatic β cells, glycolytic oscillations are in synchrony with calcium and mitochondrial oscillations to drive pulsatile insulin release, which is pivotal for the liver to regulate blood glucose dynamics. In addition, oscillation can keep biological time, essential for embryonic development in promoting cell diversity and pattern formation. The functional importance of oscillatory processes requires a re-thinking of the traditional doctrine of *homeostasis*, holding that physiological quantities are maintained at constant equilibrium values, a view that has largely failed us in the clinic. A more dynamic approach will enable us to view health and disease through a new light and initiate a paradigm shift in treating diseases, including depression and cancer. This modern synthesis also takes a deeper look into the mechanisms that create, sustain and abolish oscillatory processes, which requires the language of nonlinear dynamics, well beyond the linearization techniques of equilibrium control theory.

# Table of Contents

## 1. Homeostasis vs. oscillation in physiology

According to the doctrine of homeostasis, physiological regulation consists in maintaining key variables at constant equilibrium values, typically by using negative feedback loops. This view, influenced by control theory, leads to statements like

- 'normal body temperature is 37 degrees', maintained by a "thermostat" in the hypothalamus;
- hormones are maintained at constant levels by negative feedback involving the hypothalamus and pituitary;
- blood glucose is held at constant levels by negative feedback involving insulin secretion by the pancreas;
- gene and protein expression, for example the tumor suppressor p53, when activated, is maintained at constant levels by negative feedback from inhibitors like Mdm2.

But, while physiological regulation is clearly present in these systems, it is simply not true that physiological quantities are regulated to equilibrium values (**Fig. 1A-D**). Instead,

- core body temperature oscillates with an amplitude of ~1°C in humans (Aschoff et al., 1971) and up to 3°C in mice (Griffis et al., 2022);
- hormone levels oscillate at a number of distinct time scales in humans (Licinio et al., 1998);
- glucose and insulin concentrations in the bloodstream oscillate over time scales of 2-10 minutes (high-frequency) and 100-120 minutes (ultradian) in humans, driven by negative feedback loops containing inherent time delays (Shapiro et al., 1988; Sturis, Polonsky, Blackman, et al., 1991; Sturis, Polonsky, Mosekilde, et al., 1991);
- upon DNA damage, p53 protein levels oscillate over a 5–6-hour period, due to negative feedback by Mdm2 (Lahav et al., 2004).

Oscillation even marks the beginning of life: the first event of oocyte fertilization is the onset of intracellular calcium oscillations (Swann et al., 2006) (**Fig 1E**).

### The case of glycolysis

Oscillatory behavior in glycolysis has been widely studied since the 1960s, demonstrated by sustained oscillations in the concentrations of glycolytic intermediates such as fructose 1,6-biphosphate (FBP), ATP and NADH (Goldbeter & Berridge, 2010; Merrins et al., 2016) with a period of ~5 minutes. Glycolytic oscillations were first observed in yeast (Chance et al., 1964; Duysens & Amesz, 1957; Ghosh & Chance, 1964), but were also demonstrated in multiple mammalian systems (Frenkel, 1965; Smolen, 1995; J.-H. Yang et al., 2008).

A

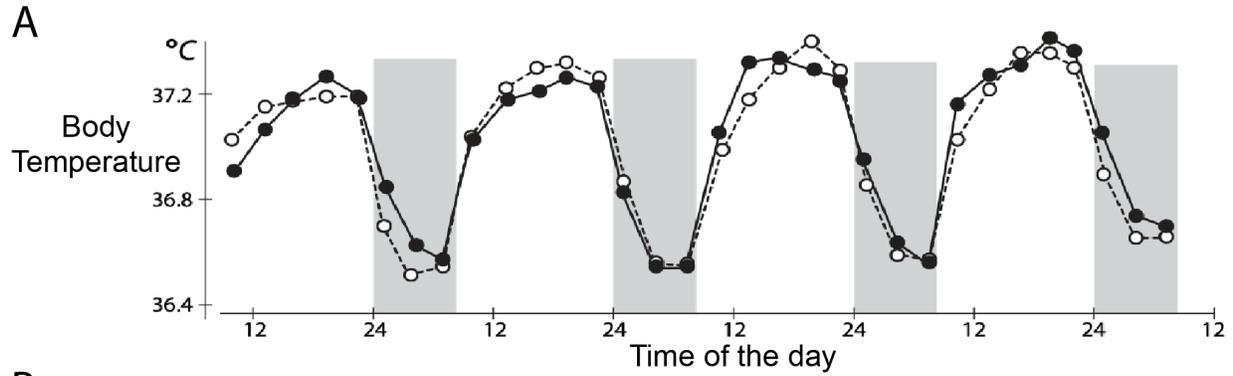

Body
Temperature

B

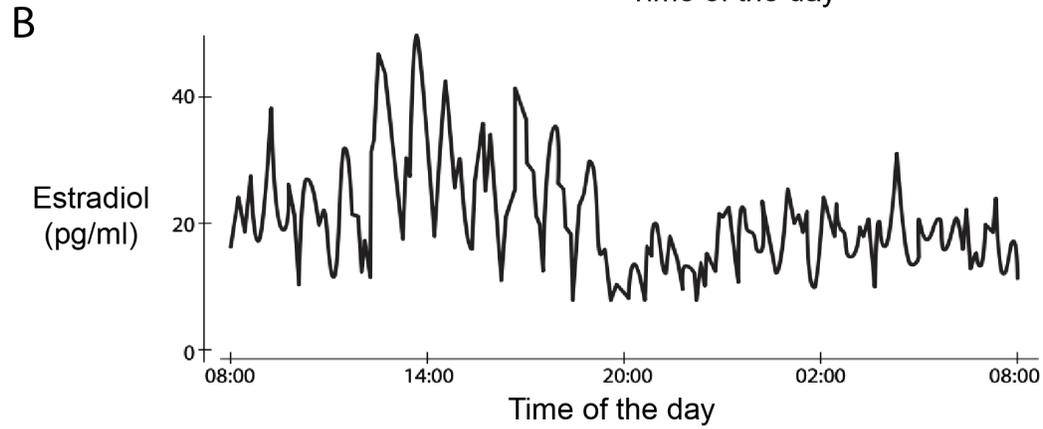

Estradiol
(pg/ml)

C

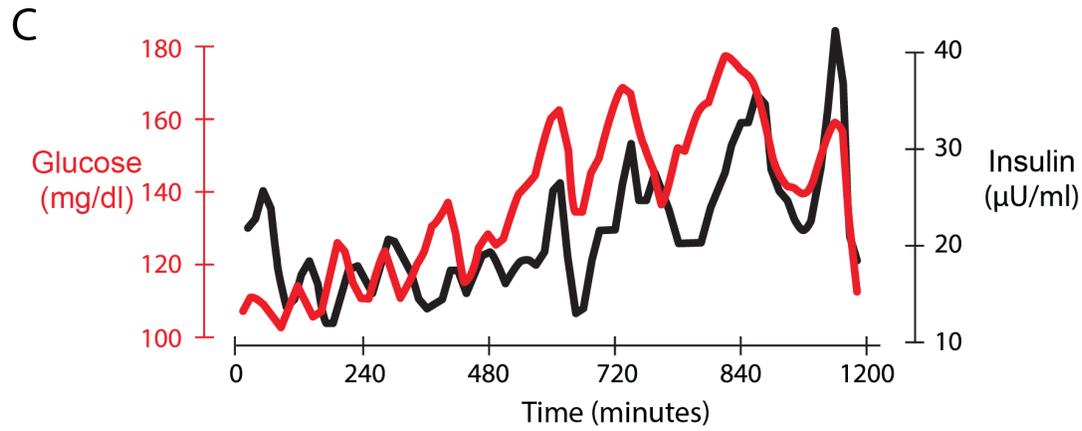

Glucose
(mg/dl)

Insulin
(µU/ml)

D

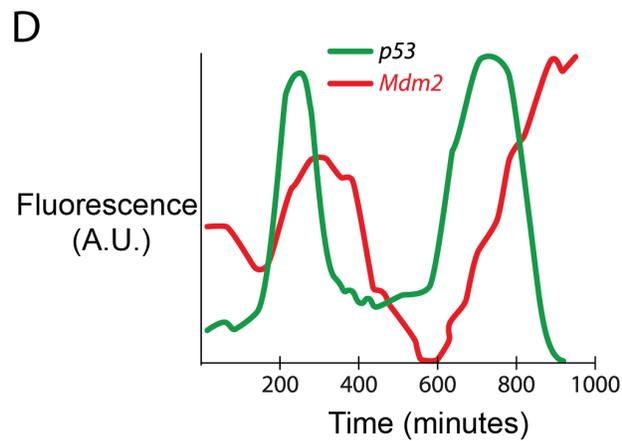

Fluorescence
(A.U.)

E

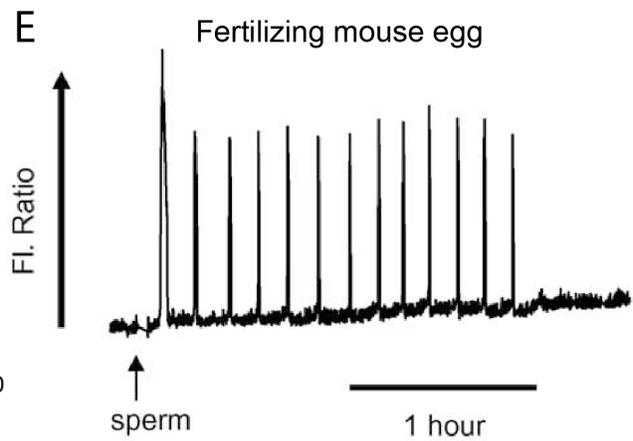

**Figure 1. Reported oscillatory behaviors in physiology.** (A) Average core body temperature (measured rectally) in six human subjects over four days (Aschoff et al., 1971). Closed circles represent the condition of an artificial light-dark cycle, while the open circles represent the same individuals in continuous darkness. Shaded areas are sleep times. (B) Multi-frequency oscillations in estradiol in a 25-year-old healthy female, at mid-to-late follicular phase (Licinio et al., 1998). (C) Glucose and insulin oscillations in a human volunteer under constant glucose infusion (Sturis et al., 1991). (D) The protein levels of the tumor suppressor p53 and its inhibitor Mdm2 show persistent oscillations in response to irradiation (Lahav et al., 2004). Figures in (A-D) were redrawn from the original publications and reprinted from "Modeling Life" by Garfinkel et al., (2017), with permission of Springer. (E) Intracellular Ca2+ oscillations induced by fertilization in a mouse oocyte (Swann et al., 2006). Reprinted from "PLCζ(zeta): a sperm protein that triggers Ca2+ oscillations and egg activation in mammals" by Swann et al., (2006), Seminars in Cell & Developmental Biology 17(2):264-73, with permission of Elsevier.

## 2. The mathematics of homeostasis and oscillation
### 2.1 Homeostasis

The mathematical expression of the doctrine of homeostasis is a stable equilibrium point (the "set point") of a dynamical system. If the system deviates from the equilibrium point, negative feedback loops bring it back. In dynamical systems theory, this behavior can be studied by linearizing the system around the equilibrium point (Hartman-Grobman Theorem) and using eigenvalues to determine the system's stability, turning the problem into a linear differential equation with the set point as its stable equilibrium point.

### 2.2 Oscillation

Oscillatory systems, by contrast, are described not by stable equilibrium points but by stable oscillations, that is, by limit cycle attractors (Garfinkel et al., 2017; Strogatz, 2015). Since there is no stable equilibrium point, and oscillation is the preferred behavior of the system, then the question of the system's regulation becomes mathematically more complex: linearization is insufficient, and nonlinear dynamics is required. In particular, if the system's behavior is described by a limit cycle attractor, then the regulatory mechanism has the job of not only creating the oscillation, but also maintaining its stability, that is, the ability of the system to return to the limit cycle after perturbation.

Nonlinear dynamics also gives us insights into the mechanisms that can create and abolish oscillations, in the qualitative changes called *bifurcations.* There are four oscillatory bifurcations that can be produced by varying a single parameter. The best-known is Hopf bifurcation, but homoclinic bifurcation, saddle-node bifurcation of cycles (or 'periodics') and saddle-node on an invariant cycle (or 'infinite-period') bifurcation also have important examples (Strogatz, 2015, Section 8.4).

Each of these bifurcations has a distinct phenomenology (**Fig. 2**), which can be used to identify it in a given system, physical or mathematical. Any one of the four bifurcations can govern the onset of an oscillation, and any one of the four can govern the offset of an oscillation, and it need not be the same bifurcation that created the oscillation. For example, one common scenario is for high-frequency oscillations to be created by Hopf bifurcation and to be extinguished by homoclinic bifurcation, a scenario that has been observed in neurons (Del Negro et al., 1998) and cardiac myocytes (Tran et al., 2009).

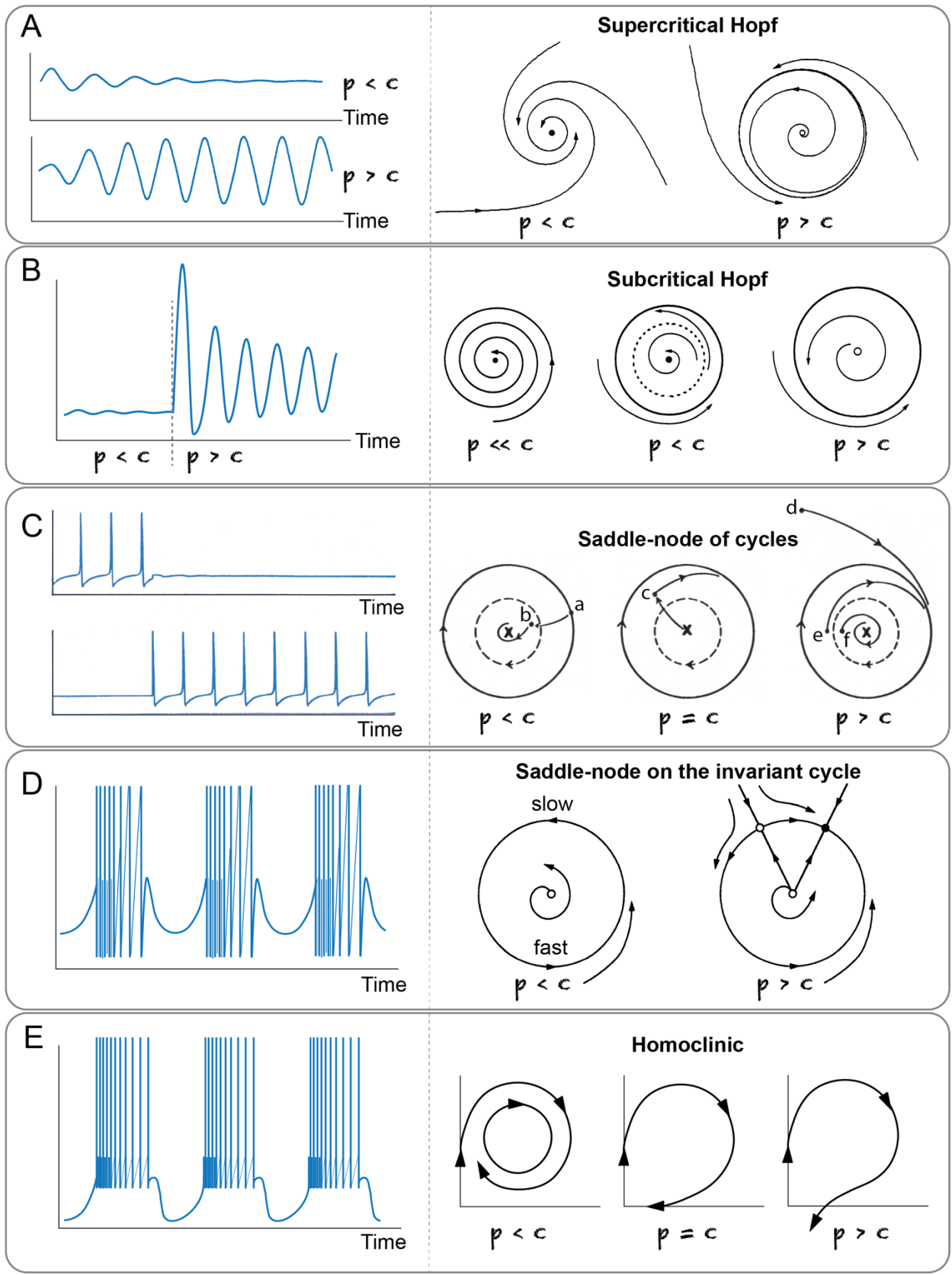

**Figure 2**. **Characteristics of bifurcation to create and destroy biological oscillations.** Left: time series. Right: 2D phase portraits (state spaces with inscribed trajectories) pre- and post-bifurcation. (A) Supercritical Hopf bifurcation. (B) Subcritical Hopf bifurcation. (C) Saddle-node of cycles. (D) Saddle-node on the invariant cycle bifurcation. (E) Homoclinic Bifurcation. See text for discussion.

In a **Hopf bifurcation**, whether supercritical (**Fig. 2A**) or subcritical (**Fig. 2B**), a stable equilibrium point becomes unstable, and a stable limit cycle appears, surrounding that now-unstable equilibrium point. In Hopf bifurcation, we should therefore see an oscillation grow gradually from zero amplitude (if the bifurcation is supercritical) around the former equilibrium point, which remains roughly in the middle of the oscillatory maximum and minimum. If the bifurcation is subcritical, the oscillation may appear full-blown at a finite amplitude instead of growing from zero, but the oscillation still surrounds the former equilibrium point. In Hopf bifurcation, as the oscillation develops, there is no significant change in frequency. These features describe Hopf Bifurcation in its forward mode. When a Hopf bifurcation governs the loss of an oscillation, the same scenario will be seen in reverse.

In a **saddle-node of cycles bifurcation** (**Fig. 2C**), a stable equilibrium point gains a pair of concentric surrounding periodic orbits. The inner periodic orbit is unstable, and the outer one is stable. Consequently, an initial condition inside the unstable periodic orbit will spiral into the stable equilibrium, while an initial condition outside the unstable periodic orbit will spiral out (or in) to the stable periodic orbit. Saddle-node of cycles bifurcations, run in reverse, can terminate oscillations, as a stable oscillation collides with an unstable periodic orbit, and is annihilated, leaving only the stable equilibrium point to govern the system.

In the saddle-node of cycles bifurcation, as opposed to the other types of oscillatory bifurcation, the initial stable equilibrium remains stable, and a new pair of periodic orbits is created. The fact that there is still a stable equilibrium point makes it possible to annihilate the oscillation with a single well-timed perturbation (Guevara, 2003; Guttman et al., 1980). Similarly, if the system is at the stable equilibrium, a single well-timed pulse can trigger sustained periodic oscillations. These phenomena cannot happen in the other oscillatory bifurcations.

The idea of "well-timed" that is used here can only be understood by reference to the nonlinear dynamics, in terms of trajectories in state space. The right-hand side of **Fig. 2C** illustrates this. For a single pulse to trigger an oscillation, it must push the state point sufficiently "outward" as defined in state space (such as the perturbation to the point c), and for a pulse to annihilate an oscillation, the pulse must push the system sufficiently "inward" to cross the unstable periodic orbit (such as the perturbation to the point b). Pushing the state point to the north (like towards the point d), when the trajectory is at high values of the Y-coordinate, will not work.

In a **saddle-node on an invariant cycle bifurcation** (**Fig. 2D**), in forward mode in the creation of an oscillation, the system state point, before bifurcation, is at rest at a stable equilibrium point on an invariant cycle. Post bifurcation, the stable equilibrium collides with an unstable equilibrium point, also on the invariant cycle. The two annihilate each other, and a stable oscillation is born. In the reverse sequence, a pair of equilibria, one stable and one unstable, appear on an invariant cycle, and the oscillation is destroyed. The state point returns to the stable equilibrium as t → ∞, thus giving this bifurcation the alternate name "infinite-period bifurcation". As the bifurcation is approached, the amplitude of the oscillations does not change, but their period increases (Strogatz, 2015; p 265)

In a **homoclinic bifurcation** (**Fig. 2E),** pre-bifurcation, a stable limit cycle co-exists with a saddle point (an equilibrium point with one stable and one unstable direction) at the origin. As the bifurcation parameter is increased (or decreased), the stable limit cycle makes contact with the saddle point and becomes an infinite period homoclinic orbit. Post-bifurcation, the oscillation is destroyed.

The identification of distinct oscillatory bifurcations is an important source of insight into the oscillations that are called "bursting". Bursting oscillations are high-frequency oscillations, often superimposed on a more slowly oscillating baseline. Under normal conditions, bursting behavior is seen in many types of neurons (Del Negro et al., 1998), and in pancreatic β cells (Bertram et al., 1995), and, under pathological conditions (see below), in other types of neurons (Y. Yang et al., 2018) and also, pathologically, in cardiac myocytes (where they are called "early afterdepolarizations") (Tran et al., 2009). Mathematically, the mechanisms of bursting onset and offset can be any one of the four bifurcations that can produce or abolish oscillation (Del Negro et al., 1998). Knowing which bifurcation underlies a given bursting behavior enables us to devise mechanistic interventions to prevent unwanted oscillations (Madhvani et al., 2011; Tran et al., 2009).

## 3. Mechanisms of homeostasis and oscillation

### 3.1 The Mechanism of Homeostasis
The standard mechanism producing homeostasis is the action of *negative feedback loops.* The homeostatic paradigm, inspired by control theory, posits a "set point" that is the target of the control mechanism (**Fig. 3**). The current state of the system is then compared to the set point, and the controller increases the state if the state is "too low" (that is, below the set point) or decreases it if the state is "too high" (that is, above the set point). But in many cases, the actual molecular and cellular basis for these mechanisms remains unknown.

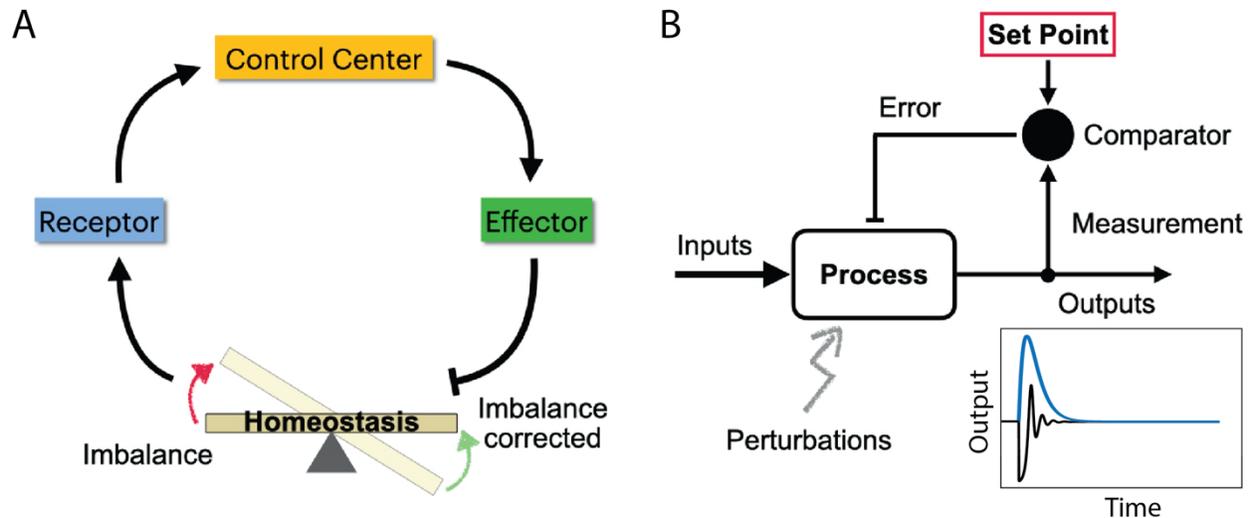

**Fig 3. The mechanism of homeostasis**. (A) Schematic of the homeostasis: negative feedback counteracts imbalances to bring the system back to its original state. (B) Control theory provides an analogy for how the control center reacts to perturbation, by 'comparing' it to a 'set point' and deriving an 'error signal' which is fed back to the system. Inset depicts responses under homeostatic control: (i) the optimal and most efficient response (blue), featuring a mono-exponential approach to equilibrium; (ii) a typical but suboptimal response (black), with rapid approach but damped oscillations before returning to equilibrium.

### 3.2 The Mechanisms of Oscillation
### 3.2.1 Mechanism #1: steep negative feedback plus time delay

While negative feedback loops can produce a static equilibrium, the same loop can also produce oscillations. The conditions under which negative feedback produces one or the other response have been studied in many models (**Fig. 4A**). The pioneering work of Mackey and Glass studied a model of the negative feedback loop that controls $CO_2$ levels by adjusting the respiratory rate (Mackey & Glass, 1977).

Using bifurcation theory, Mackey and Glass were able to show that their system would undergo a qualitative change in its behavior, a Hopf bifurcation, in which the formerly stable "set point" of the controller becomes unstable and is replaced by a stable oscillation. Their model contained a parameter $n$ that controlled the steepness of the negative feedback, and another parameter $\tau$ that reflected the time delay in the loop. They then showed that the system would undergo a Hopf bifurcation, and begin to oscillate, if the negative feedback is steep enough *and* there are sufficient time delays in the system. They derived a criterion for the Hopf bifurcation: the equilibrium point became unstable, and a stable oscillation appeared, when the product of $n$ and $\tau$ exceeded a certain quantity (**Fig. 4B**).

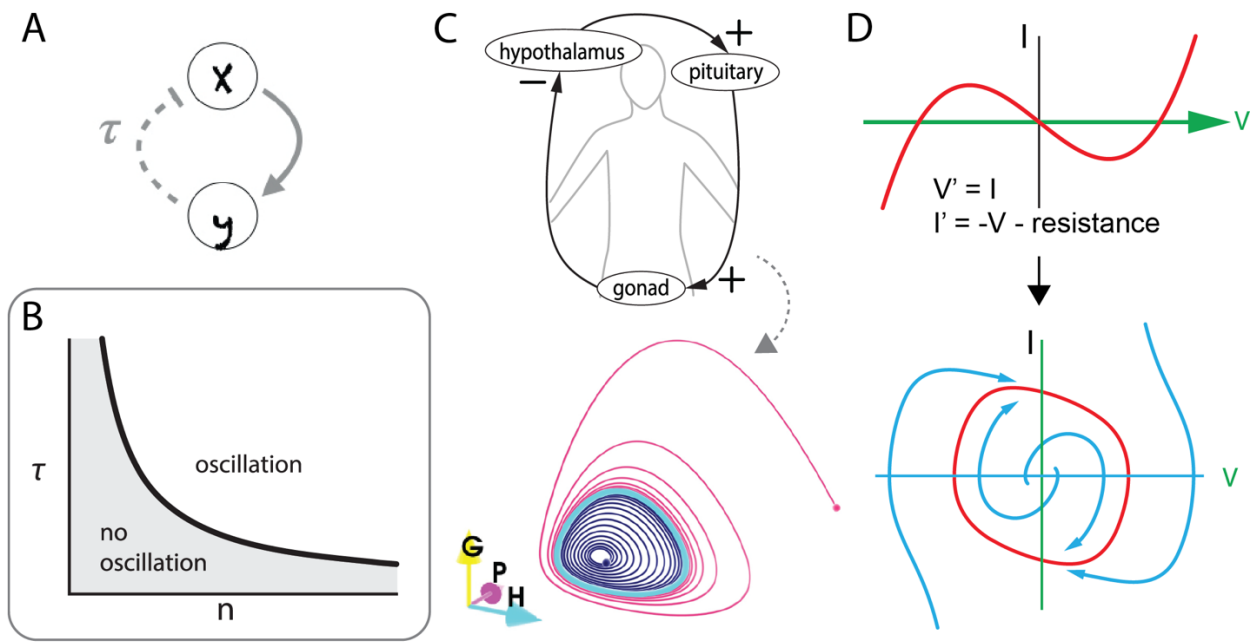

**Figure 4. Mechanisms of oscillation. (**A) Schematic of a 2-variable negative feedback loop with explicit time delay $\tau$. (B) Hopf bifurcation boundary for the negative feedback loop (n = slope of negative feedback). (C) A 3-variable model of the HPG negative feedback loop, without explicit time delays (upper), can produce oscillation (lower) if the negative feedback is sufficiently steep. (D) Schematic of a 2-variable "negative resistance" system (upper) can exhibit sustained oscillations, governed by a stable limit cycle attractor in 2D state space (lower).

Thus, nonlinear dynamics gives us a precise answer to the question: when does a negative feedback loop sustain equilibrium and when does it promote oscillation?

The combination of sensitive negative feedback plus time delays is probably the most common mechanism for biological oscillations, both good and bad. For example, the 2017 Nobel Prize for Physiology or Medicine was awarded to Hall, Rosbash and Young for the discovery of the mechanism of the circadian rhythm, in which "several transcription factors operate in a genetic network incorporating autoregulatory [negative] feedback loops. Oscillations are achieved by delaying various steps in the network. For example, accumulation of one of the transcription factors — Period — is retarded in the cytoplasm by phosphorylation and degradation" (Young & Kay, 2001).

"Negative feedback with time delays" is also the driver of many functional rhythms in endocrine systems (**Fig. 4C**). For example, gonadal hormones exert negative feedback on hypothalamic secretion of hormone releasing factors. A Hypothalamic-Pituitary-Gonadal negative feedback loop will produce stable endocrine oscillations when the negative feedback becomes sufficiently steep (Garfinkel et al., 2017; W. R. Smith, 1983). In Smith's phrase, "Puberty is a Hopf Bifurcation," meaning that the HPG system will begin cycling when the hypothalamus becomes more highly sensitive to negative feedback from the gonads. There are also oscillations driven by negative feedback loops in the Hypothalamic-Pituitary-Adrenal Cortex, producing oscillations in cortisol (Lightman et al., 2020; Walker et al., 2012).

When oscillation is produced by a "negative feedback plus time delay" mechanism, the bifurcation underlying the onset of oscillation is a Hopf bifurcation, as Mackey and Glass found in their system and Smith found in the HPG system.

### 3.2.2 Mechanism #2: "negative resistance"

There are other mechanisms of oscillation, some of which can be seen even in two-variable systems with no time delays. One simple model displaying this property is the Fitzhugh-Nagumo neuronal model (Keener & Sneyd, 2009: p 221) (see also Garfinkel et al., 2017: p 217). In this model, for all non-zero values of "negative resistance", there is short-range destabilizing positive feedback, making the equilibrium point inherently unstable (**Fig. 4D**). The negative feedback at high voltage is stabilizing and forces the state point to spiral inward. Trapped in the 2-D plane, between the spiraling-out at low voltages and the spiraling-in at high voltages, there is a single closed orbit, a (stable) limit cycle attractor (**Fig. 2A**).

This mechanism can be thought of as "short-range activation plus long-range inhibition", where 'short and 'long' refer to state space. The phrase is taken from Gierer and Meinhardt's pioneering studies on *spatial* pattern formation (Gierer & Meinhardt, 1972), where the activation and inhibition are short-range and long-range in *physical* space. This was their generalization of the original mechanism for spatial oscillations proposed by Turing (1952).

Another example of negative resistance oscillators are the membrane voltage oscillations driven by the negative resistance region of the NMDA receptor (Brodin et al., 1991), which have been found to be important in psychiatric depression (Y. Yang et al., 2018) (see also Discussion). Recent work based on this observation has highlighted a mechanism, and a mechanistic approach to therapy, for the pathological oscillations that are seen in psychiatric depression. In rodent models of depression, including chronic restraint stress, neurons in the Lateral Habenula (LHb) have emergent high-frequency bursting oscillations in membrane potential (Yang et al., 2018). This bursting behavior then facilitates "synchronization" to Local Field Potentials, another hallmark of depression. Yang et al. showed that the bursting oscillation was itself caused by the activity of the NMDA receptor in LHb neurons, whose negative-resistance region turns the neuron into a negative-resistance oscillator. Their analysis provided a mechanism for therapy: the NMDA receptor blocker (and antidepressant) ketamine abolished bursting oscillations by reducing the negative resistance region, and ameliorated depression-like behavior.

## 4. Oscillation: physiological functions

Oscillations may be everywhere in physiological systems, but their physiological function is far from clear. Circadian rhythms clearly evolved to adapt to and take advantage of the external day/night cycle, but the role of other rhythms often remains elusive. It has even been argued that many finer-scale oscillations might not serve any biological functions at all, as has been suggested for glycolysis (Alberts, 2015; Chandra et al., 2011; El-Samad, 2021).

Here, we summarize representative findings to provide evidence that oscillations are not an unwanted product of negative feedback regulation. Rather, they represent an essential design feature of nearly all physiological systems.

### 4.1 Oscillation as the preferred mode for communication
### 4.1.1 Avoiding adaptation and desensitization

In inter- and intra-cellular communication, oscillatory signaling is generally the preferred mode of behavior. One reason for this is that, in many systems, constant high levels of signal lose their effect due to adaptation and saturation at the 'receiver'. For example, only periodic cAMP signals delivered with physiological frequency (every 5 minutes) could stimulate the aggregation and differentiation of starved social amoebae (Goldbeter, 1988), because cell-surface cAMP receptors become desensitized and degraded when stimulated with constant signals (Martiel & Goldbeter, 1987; Van Haastert et al., 1992). Such inhibition of surface receptors because of chronic stimuli also applies to the receptors for insulin-like growth factor I (IGF-1) (Norstedt & Palmiter, 1984) and to the epidermal growth factor receptor (EGFR) (Klein et al., 2004).

### 4.1.2 Avoiding toxicity

Oscillatory signaling is also a solution to the dilemma that constant high levels can be toxic. In response to moderate DNA damage, p53 oscillates to trigger cell cycle arrest and DNA damage repair, a property that is essential for tumor suppression (M. S. Heltberg et al., 2022; Lahav et al., 2004; Xiong & Garfinkel, 2022). Cells with constant p53 of similar amplitude, on the other hand, readily undergo senescence or apoptosis within hours (Purvis et al., 2012). Moreover, constant high levels of p53 are known to be embryonic lethal (Marine et al., 2006) and can cause widespread tissue damage (Moyer et al., 2020). A similar phenomenon happens in the immune response: NF-kB activity in the nucleus oscillates when stimulated by endogenous or mild immunogenic signals, such as TNF-alpha and poly(I:C) (Adelaja et al., 2021), to selectively induce target gene expression (M. L. Heltberg et al., 2019; Hoffmann et al., 2002). Evidence also has suggested that this oscillatory behavior could avoid extensive epigenome remodeling that is costly, as in extreme immune challenges, or deleterious, as in autoimmune diseases (Cheng et al., 2021).

### 4.1.3 Resistance to corruption by noise

Another desirable feature of oscillation is that oscillatory signaling is more resilient to noise and more resistant to corruption of its information. Information encoded in the frequency (i.e., frequency modulation [FM]) is less sensitive to corruption by noise than information encoded in the amplitude (i.e., amplitude modulation [AM]). This is why FM radio is superior to AM in sound quality. The transmission of biological information seems to follow the same principle, exemplified by the activation of Ca2+/calmodulin-dependent protein kinase II (CaMKII), an enzymatic complex consisting of dimers of hexameric rings. CaMKII is highly sensitive to the frequency of intracellular Ca2+ oscillations – only high frequency enables consecutive and autonomous activation of individual catalytic domains – but not to pulse duration or amplitude (De Koninck & Schulman, 1998).

## 4.2 Oscillations reconcile incompatible biological processes

As early as 1980, Boiteaux, Hess, and Sel'kov suggested that oscillation is the preferred mode of operation when it is necessary to reconcile between incompatible conditions (Boiteux et al., 1980). Like the sleep-wake cycle in the circadian rhythm, or the longer-term hibernation cycle through seasons, many fine-scale biological oscillations serve to accommodate processes that cannot occur simultaneously.

### 4.2.1 Cell cycle

The cell cycle progresses through distinct phases to achieve cell division and proliferation: DNA synthesis, cell growth, mitosis, and various checkpoints (Alberts, 2015). Obviously, DNA synthesis cannot happen during mitosis, during which the chromatin condenses, and DNA become inaccessible. Therefore, the cell cycle must accommodate incompatible conditions, by cycling among them. The engine driving the cell cycle oscillation is the negative feedback loop of cyclin and cdc2/APC, acting with time delays (Ferrell et al., 2011; Tyson, 1991). The negative feedback loop serves as the mechanism to enforce the overall strategy of reconciling these incompatible processes through oscillation.

### 4.2.2 Yeast metabolic cycle

Another excellent example is the yeast metabolic cycle, as studied by Tu et al. (2005). Their research uncovered a temporal profile of gene expression in yeast cultures that was highly periodic, synchronized to the respiratory oscillation (~5 hours). Even more remarkably, the overall cycle was clearly divided into 3 distinct phases, with a closely coordinated group of genes being expressed in each phase. For example, genes for amino acid synthesis were expressed during the "Oxidative" phase, and genes for mitochondrial processes were expressed during the "Reductive/Building" phase, while genes for heat shock proteins were expressed during the "Reductive/charging" phase. The authors referred to this time-sharing arrangement as "temporal compartmentalization", a concept that was echoed by a synthetic design of a gene-metabolic oscillator in *E. coli* (Fung et al., 2005). Ultradian gene expression was later found in mammalian systems, in both nocturnal (Hughes et al., 2009) and diurnal (Mure et al., 2018) animals.

### 4.2.3 Energetic oscillations in mitochondria

At a finer timescale, mitochondrial function is characterized by oscillation in redox state, alternating between an oxidative and a reduced environment. In guinea pig cardiomyocytes, mitochondrial redox transitions were shown to oscillate with a period of 1-3 minutes (O'Rourke et al., 1994; Romashko et al., 1998), as measured by the fluorescence of flavoproteins, which is closely linked to mitochondrial NADH levels.

Oxidative processes (such as the TCA cycle) accumulate NADH, increasing fluorescence levels; while reductive processes (including the electron transport chain) use NADH, diminishing fluorescence. The oxidative/reductive oscillation is closely linked to oscillation in mitochondrial membrane potential ($\Delta\Psi_m$), which is a 2-stage alternation, staggering the build-up of proton gradient in one phase from mitochondrial ATP synthesis (by draining the proton gradient) in the other (Aon et al., 2003; Romashko et al., 1998).

### 4.2.4 Proliferation versus commitment in development

Oscillatory behaviors also permeate developmental processes, where alternation between proliferation and differentiation is required (Beets et al., 2013). In developing vasculature, for example, proliferation and differentiation alternate over a 24-hour cycle to achieve morphogenesis (Guihard et al., 2020). The clock driving the cycle is the negative feedback loop between BMPs 4 and 9 and their inhibitors MGP and CV2, respectively. Each cycle has a proliferative phase of BMP4 dominance-BMP9 inhibition, and a differentiation and commitment phase of BMP4 inhibition-BMP 9 dominance. Another example of developmental oscillations is the ERK oscillation that is downstream of EGF/FGF stimulation and MAPK pathway activation. The temporal dynamics of ERK are known to govern cellular proliferation and differentiation (Marshall, 1995; York et al., 1998), not only seen in embryonic development (Kholodenko et al., 2010; Raina et al., 2022), but also in tissue regeneration (De Simone et al., 2021) and tumor microenvironment (Davies et al., 2020; Gillies et al., 2020). Oscillations in Wnt and Notch signaling could play a similar role (Sonnen et al., 2018).

### 4.3 Oscillations allow for synchronization of coupled biological processes

One of the most important functions of oscillation in biology is that systems that are time-varying can lend themselves to synchronization of their time-varying processes, thereby combining many small outputs into one large one. This is seen in the cellular slime mold *D. dictyostelium*: individual cells synchronize their pulsations of cAMP to achieve macroscopic quantities of the chemoattractant, which produces aggregation (reviewed in Goldbeter and Berridge, 2010). In a striking example of the benefits of oscillation, cell-wide mitochondria were also found to synchronize their oscillatory metabolic activities for maximum ATP output in cardiac tissues (Aon et al., 2003, 2006).

Similarly, in all secretory organs of the body, including the pituitary and pancreas, cells secreting a given hormone must synchronize their microscopic pulsations to achieve the macroscopic output necessary for physiological function. In response to elevated glucose levels in the blood, the pancreas secretes insulin in an oscillatory fashion, with an amplitude of up to 600 pmol/L and a period of ~5 minutes, a phenomenon that has been observed in rodents, canines and humans (Lang et al., 1979; Matveyenko et al., 2008). The liver's response to insulin seems to depend on this high-frequency oscillation of insulin release into the blood; its disruption could contribute to type 2 diabetes (Satin et al., 2015).

### 4.3.1 The intracellular oscillator in the pancreatic β cell

Several important studies have elucidated the intracellular mechanisms of oscillatory insulin release in the pancreatic β cell (McKenna et al., 2016; Merrins et al., 2016), highlighting the pivotal role of glycolytic oscillations (O'Rourke et al., 1994). In particular, a rise in glycolytic activity has been shown to precede membrane depolarization and calcium influx, which in turn triggers oxidative phosphorylation and ATP depletion (**Fig. 5**). Some authors have even proposed that "pulsatile basal insulin secretion is driven by glycolytic oscillations" (Fletcher et al., 2022) and metabolic cycles (Merrins et al., 2022). Below, we summarize these findings for how glycolytic oscillations can drive pulsatile insulin release, by coupling to metabolic oscillations in the mitochondria and to oscillations in intracellular $Ca^{2+}$.

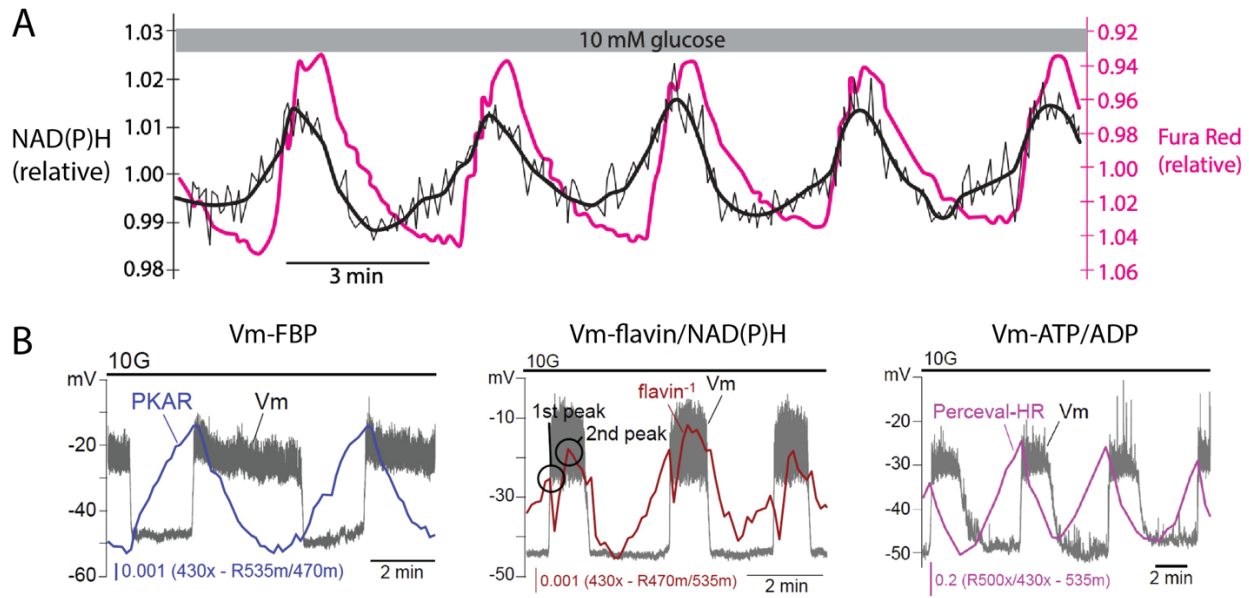

**Figure 5. Intracellular processes are coupled in pancreatic β cells to release insulin in an oscillatory fashion.** (A) Metabolic oscillations (NAD(P)H) are synchronized to oscillations in intracellular $Ca^{2+}$ (Fura Red) in mouse pancreatic β cells. Redrawn from "$Ca^{2+}$ controls slow NAD(P)H oscillations in glucose-stimulated mouse pancreatic islets" by D.S. Luciani, S. Misler, and K.S. Polonsky, (2006), Journal of Physiology 572(2):379–392, and reprinted from "Modeling Life" by Garfinkel et al., (2017), with permission of Springer. (B) Coupling of metabolic intermediates to membrane potential (Vm) oscillations in mouse pancreatic β cells. Example recordings of Vm-FBP (left), Vm-Flavin/NAD(P)H (middle), and Vm-ATP/ADP ratio (right). Reprinted from "Phase analysis of metabolic oscillations and membrane potential in pancreatic islet β-cells" by Merrins et al., (2016), Biophysical Journal 110:691-699, with permission of Elsevier.

The coupling of glycolytic oscillations to mitochondrial and intracellular $Ca^{2+}$ oscillations cycle through the following four stages to secrete insulin in a pulsatile fashion (**Fig. 6**):

1. As glucose is imported into the cell, it feeds into glycolysis in the cytosol, producing ATP and pyruvate. In the mitochondria, pyruvate propels the TCA cycle, generating NADH for charging the electron transport chain, which pumps H+ into the intermembrane space, thus increasing mitochondrial membrane potential ($\Delta\Psi_m$) (Merrins et al., 2016).

2. As glycolysis continues, ATP accumulates. When the ATP/ADP ratio reaches a critical value, the ATP-sensitive K+ channels ($K_{ATP}$ channels) on the cell membrane become inhibited, causing membrane depolarization (Craig et al., 2008). In turn, this triggers $Ca^{2+}$ influx through the voltage-dependent $Ca^{2+}$ channel (Satin, 2000).

3. Influx of $Ca^{2+}$ then plays numerous roles inside the cell, including ramping up mitochondrial ATP production (Jouaville et al., 1999) and triggering insulin release through exocytosis (Kesavan et al., 2007), a process that rapidly consumes ATP.

4. Decreased ATP/ADP ratio then re-activates the $K_{ATP}$ channels, therefore restoring the membrane potential and stalling $Ca^{2+}$ influx, which allows for glycolysis to build up ATP again.

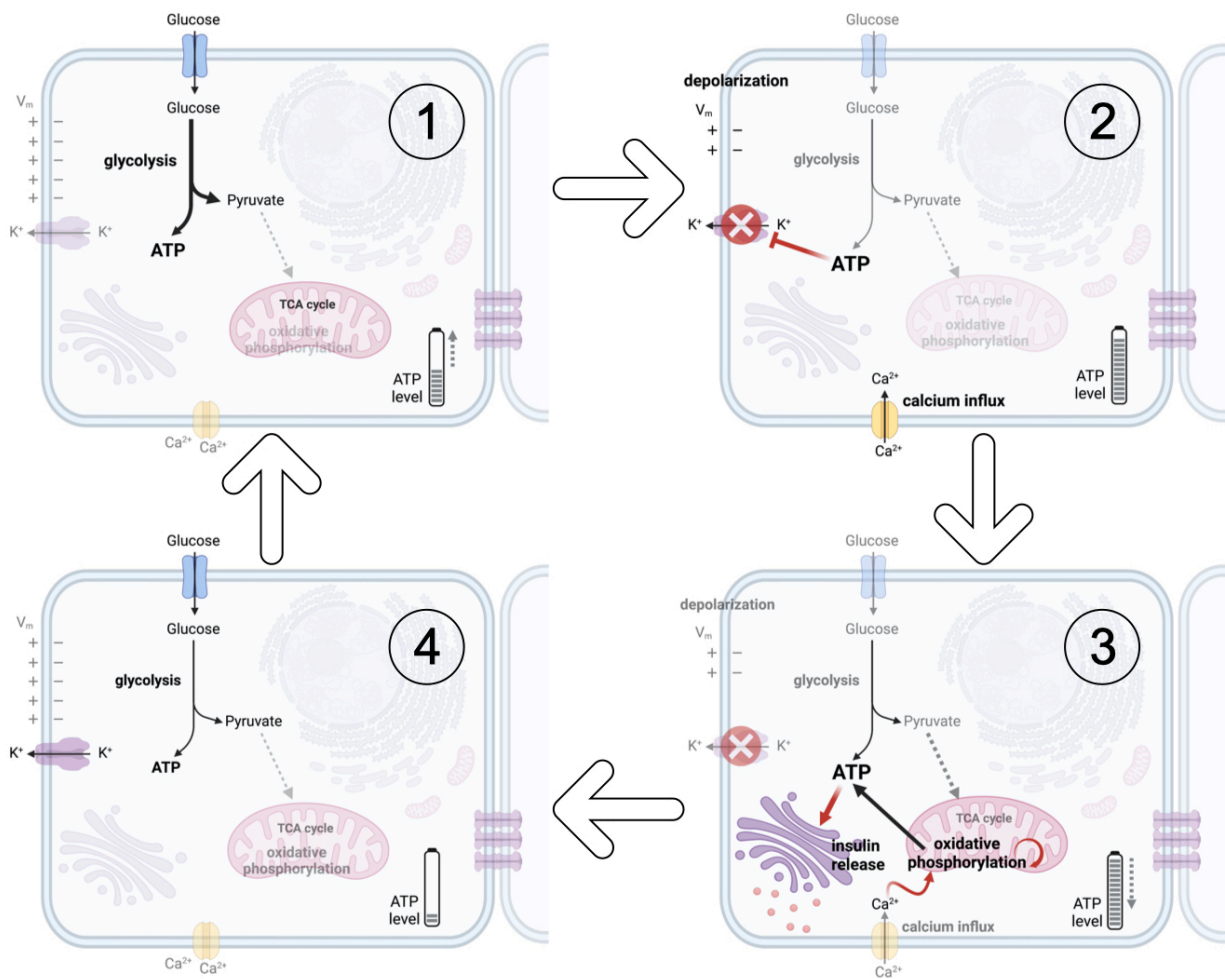

**Figure 6. The four stages of oscillatory insulin release in pancreatic β cells (see text).**

### 4.3.2 Tissue-level synchronization

These cellular oscillators must then be coupled into a synchronous macroscopic oscillation, to achieve their physiological effects. The process by which this synchronization takes place has been well studied: pancreatic β cells within the islets of Langerhans are connected by gap junctions (**Fig. 6**), which creates synchronous insulin release within the islet (Benninger et al., 2011; Sherman et al., 1988). On a higher level, inter-islet synchronization could be guided by the shared glucose-rich environment (Bier et al., 2000) and/or neuronal input from the liver (Imai et al., 2008).

### 4.3.3 Oscillatory processes across different organ systems

At still higher levels of organization, oscillatory processes must often be regulated and synchronized from organ to organ. As mentioned above, many hormones released by the pituitary show ultradian rhythms in humans, including luteinizing hormone (LH) (Bäckström et al., 1982), follicle-stimulating hormone (FSH) (Bäckström et al., 1982), adrenocorticotropic hormone (ACTH) (Brandenberger et al., 1987), thyroid-stimulating hormone (TSH) (Romijn et al., 1990), growth hormone (Goji, 1993), and prolactin (Bäckström et al., 1982; Veldhuis & Johnson, 1988). The role of the hypothalamus-pituitary-gonad (HPG), -thyroid (HPT), and -adrenal (HPA) axes in regulating the oscillatory release of these hormones has been well-characterized (Grant et al., 2018). Studies have also revealed that the ultradian oscillation in leptin, an adipocyte hormone (Sinha et al., 1996), showed pattern synchrony with those of LH and estradiol in healthy women (Licinio et al., 1998), and with those of TSH in healthy adults (Mantzoros et al., 2001), but is inversely related to ACTH (Licinio et al., 1997). These synchronized actions could represent a general strategy for organ-organ communication.

And, at the highest level, since virtually all organs show rhythmic behaviors that are autonomous to that organ, there is clearly a need to regulate these peripheral oscillations to the overall circadian cycle, a need for "one ring to rule them all…and in the darkness bind them". It has been suggested that oscillations in core body temperature could serve as this universal entrainer (Brown et al., 2002; Buhr et al., 2010; Qian et al., 2022).

### 4.4 Oscillations promote cell diversity and pattern formation

Early studies demonstrated how biodiversity can be achieved by species oscillations (Huisman & Weissing, 1999; Kerr et al., 2002). Population oscillations make it possible for antagonistic organisms to share the same space, by cycling their populations in time, as demonstrated in rock-scissors-paper models (Elowitz & Leibler, 2000). This notion turned out to have wide implications beyond ecology. It facilitates biodiversity by enabling time-sharing, or 'temporal compartmentalization', in the phrase of Tu et al. (2005).

The segmentation clock in mammalian embryonic development has been studied since the 1990s (Lewis, 2003; Monk, 2003; Palmeirim et al., 1997), where the highly conserved oscillatory behavior in Hes1 protein level is known to keep biological time for coordinated somite formation (Diaz-Cuadros et al., 2020; Matsuda et al., 2020). Hes1 represses downstream target Dll1, a ligand for the Notch signaling pathway (Kobayashi et al., 2009). Notch signaling in embryonic stem cells favors neural differentiation, whereas its inactivation tilts towards cardiac mesoderm differentiation. Here, Hes1 oscillations could generate cell populations that are heterogeneous in their differentiation competency, thus creating diverse cell lineages for somitogenesis (Kobayashi & Kageyama, 2010). Through ligand-receptor interactions, the segmentation clock also gives rise to spatial organization of diverse cell populations during somitogenesis and other processes of pattern formation (Bocci et al., 2020; Oates et al., 2012).

## 5. Discussion

Equilibrium-centric dynamics and homeostasis have long dominated biological thinking. However, we are now learning that this view is far from complete. Here, we studied examples of physiological oscillation and summarized their mechanisms and the mathematics required to understand them. Importantly, these oscillations play key functional roles, by being the preferred mode of communication in cell signaling, reconciling incompatible processes, synchronizing coupled processes, keeping biological time, and promoting diversity and pattern formation (**Fig. 7**).

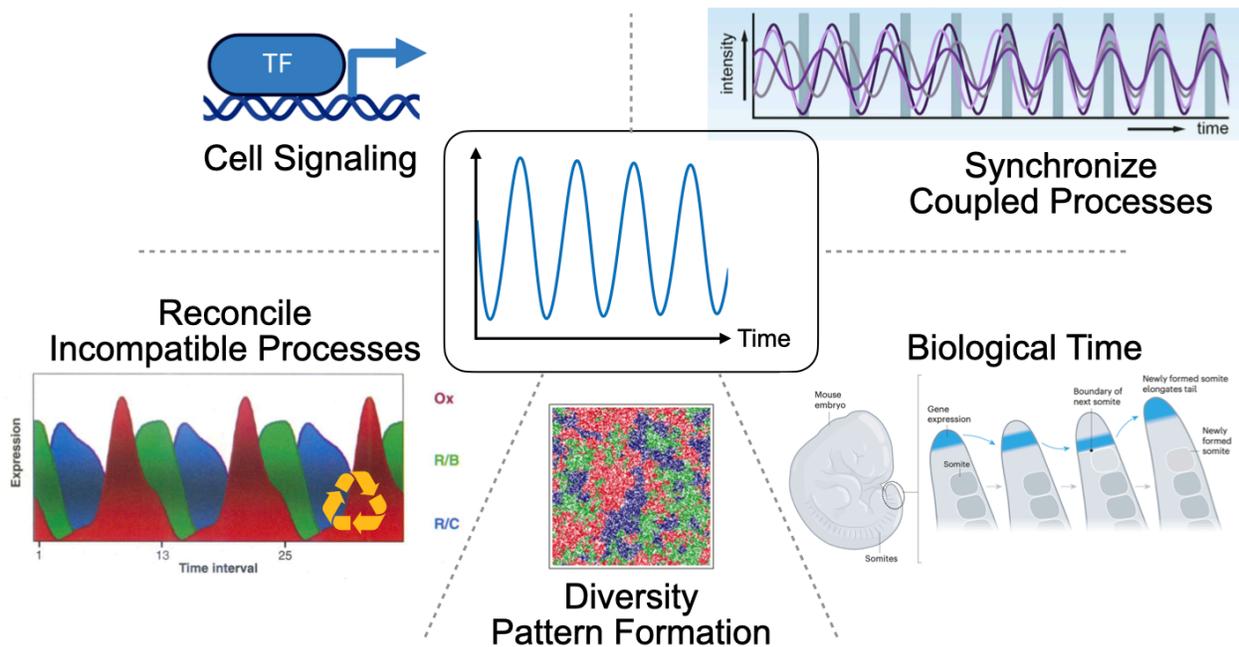

**Figure 7. Physiological functions of biological oscillations.**

As new oscillatory behaviors are being discovered in physiological systems, additional functions may emerge. For example, high frequency voltage oscillations, as recorded by contact electrodes in the cortex, have been suggested to be critical in intracortical communication in the brain (Engel et al., 2009). It has been proposed that these oscillations can synchronize neuronal activity over long distances, thereby providing a basis for how different regions of the brain, say auditory and visual, can combine ("bind") their information into a single object (Buzsáki, 2006). Researchers have found that these high-frequency oscillations ("ripples") phase-synchronize, with definite phase-offsets, over long distances in the brain. This phase-synchrony, they suggest "may help to "bind" different aspects of a mental event encoded in widespread cortical areas into a coherent representation" (Dickey et al., 2022).

## 5.1 Beyond homeostasis: homeodynamics?

If oscillatory processes are central to physiology, then we will need to take a fresh look at the doctrine of homeostasis. We suggest that the concept of homeostasis needs to be parsed into two separate components. The first component is the idea that physiological processes are regulated and must respond to environmental changes. This is obviously critical and true. The second component is that this physiological regulation takes the form of control to a static equilibrium point, a view which we believe is largely mistaken. "Homeodynamics" is a term that has been used in the past to try to combine regulation with the idea that physiological processes are oscillatory (Lloyd et al., 2001; Soodak & Iberall, 1978; Yates, 1994). It may be time to revive this terminology.

In past decades, the idea of homeostasis has been tacitly supported by forms of data acquisition that consisted in taking snapshots of cell activity, such as in single-cell RNA-seq experiments (Lähnemann et al., 2020). With new data modalities such as single-cell live imaging of gene expression (Phillips et al., 2017) and time-series RNA-seq (L. M. Smith et al., 2020) as well as emerging computational methods like Oscope (Leng et al., 2015), we will soon be able to detect oscillatory behaviors in a high-throughput manner, to survey dynamic processes in biology systematically.

## 5.2 Oscillation in health and disease

The dynamic paradigm offers a new approach to our concepts of health and disease. In homeostasis, "levels" are the goal of the system and the target of therapies. Health is having the right levels, and disease is marked by levels that are either too low or too high. Therapy consists of supplying agonists (if levels are too low) or antagonists (if levels are too high). By contrast, the dynamical conception of health looks at the body's oscillatory processes and asks if they are in the right dynamical ranges, and whether they are properly synchronized (and anti-synchronized) to other oscillatory processes. In this view, disease can manifest as a loss of oscillation (Knobil et al., 1980; Xiong & Garfinkel, 2022) and/or a loss of synchronization (Glass, 2001; Goldbeter, 2002; Noble, 2006; Tu et al., 2005).

An excellent example of the conflict between the homeostatic and dynamic views of health and disease can be found in recent thinking about psychiatric depression. The homeostatic focus on 'levels', when applied to depression, led to the serotonin hypothesis, which held that depression was caused by low levels of serotonin, and that Selective Serotonin Reuptake Inhibitors (SSRIs) would restore this "imbalance".

But recent work has cast doubt on the 'serotonin levels' hypothesis. A recent exhaustive meta-analysis of serotonin studies in depression concluded that "The main areas of serotonin research provide no consistent evidence of there being an association between serotonin and depression, and no support for the hypothesis that depression is caused by lowered serotonin activity or concentrations". *The New York Times*, reporting this study, noted that "The most commonly prescribed medications for depression are somewhat effective — but not because they correct a "chemical imbalance" (Moncrieff et al., 2022; D. G. Smith, 2022).

In contrast with this static view, we saw above a radically different picture of depression as caused by a synchronization of bursting oscillations in a certain brain region. The mechanistically derived therapy for this was to abolish the bursting oscillations by blocking the negative-slope resistance region of the neuronal NMDA receptor.

## 5.3 Bad Vibrations

As we have already seen in the case of depression, not all oscillations serve a positive physiological function. Oscillations in physiology can also be pathological and dysfunctional. One important example of 'bad oscillations' is seen in the onset of seizure, which presents in the EEG as large well-formed oscillatory behavior (Marten et al., 2009). Bad oscillations also include the onset of muscle tremor in Parkinsonism, stroke, and demyelinating diseases. In the latter two cases, the onset of oscillation can be attributed to the destabilization of the negative feedback stretch reflex, which normally maintains positional homeostasis in the limbs. Stroke patients suffer from 'hyperreflexia', increasing the steepness of the negative feedback loop, and patients with demyelinating diseases like MS suffer from increased time delays in those feedback loops, due to lowered axonal conduction velocity (Garfinkel et al., 2017).

## 5.4 Glycolysis Revisited

Oscillations in glycolysis have recently become a point of conflict between the homeostatic and the dynamic worldview. The controversy is about whether these oscillations play a physiological role. A number of authorities have suggested that oscillations in glycolysis are merely an accidental and unwanted side effect of homeostatic negative feedback regulation: 'oscillations become inevitable' (Alberts, 2015), simply because there are many negative feedback loops (Chandra et al., 2011; El-Samad, 2021).

The paradigm of homeostasis, when applied to glycolysis, assumes that the purpose of glycolysis is to maintain constant ATP levels (Chandra et al., 2011). And yet all the available data suggest that glycolysis is oscillatory, and *needs* to be oscillatory, in order to couple to mitochondrial oscillations and to oscillations in intracellular $Ca^{2+}$. Indeed, as we saw, it has even been suggested that glycolytic oscillations are the engine driving pulsatile insulin secretion in the pancreas (Section 4.3.1).

In the face of all the data indicating both the fact of oscillatory glycolysis and its functional role, the idea that glycolysis *must* be maintaining constant ATP levels comes therefore not from data, but from a philosophical belief, that control *means* equilibrium control. The virtues of oscillation do not appeal to this control-theory mindset. Rather, oscillation is viewed as highly undesirable, and is disparaged as "fragility" (seven times in Chandra et al., 2011). But oscillation, as we have seen, is an important functional mode of operation in physiological systems. And oscillatory behaviors are governed by limit cycle attractors that are stable and robust to perturbation, making them responsive, and not at all "fragile". Indeed, recent work has suggested that glycolytic oscillations are "thermodynamically optimal" (Kim & Hyeon, 2021). If the purpose of glycolytic regulation is in fact to produce a (regulated) *oscillatory* response, then glycolytic oscillations may not be just an inevitable side effect of the feedback loops. Instead, they could well be the design principle of glycolytic physiology (Goldbeter, 2018).

## 5.5 Outlook

It is now 50 years since the appearance of P.W. Anderson's visionary "More is different: Broken symmetry and the nature of the hierarchical structure of science" (Anderson, 1972; Stumpf, 2022). Among the broken symmetries that Anderson discussed is the emergence of "oscillators", which violate time-invariance symmetry. He asserts that "Temporal regularity [i.e., oscillation] is very commonly observed in living objects", and that the role of this "regular pulsing" is, first, in "extracting energy from the environment in order to set up a continuing, quasi-stable process"; second, as a means of "handling information"; and third, to use "phase relationships of temporal pulses" to regulate physiological processes. It is time to bring this conception of physiological oscillations to the center of biological discourse.